\documentclass[12pt]{iopart}
\usepackage{iopams}  
\usepackage{graphicx}

\begin{document}

\title[ ]{Entanglement, measurement, and conditional evolution of the Kondo 
singlet interacting with a mesoscopic detector}

\author{Kicheon Kang}
\address{Department of Physics, Chonnam National University, 
 Gwang-Ju 500-757, South Korea}
\ead{kckang@chonnam.ac.kr}
\author{Gyong Luck Khym}
\address{Department of Physics, Chonnam National University, 
 Gwang-Ju 500-757, South Korea}
\ead{glgreen@boltzmann.chonnam.ac.kr}
\begin{abstract}
We investigate various aspects of the Kondo singlet in a quantum dot (QD)
electrostatically coupled to a mesoscopic detector. 
The two subsystems are represented by an entangled state between the
Kondo singlet and the charge-dependent detector state.
We show that the phase-coherence of the Kondo singlet
is destroyed in a way that is sensitive to the charge-state information
restored both in the magnitude and in the phase of the scattering coefficients
of the detector. 
We also introduce the notion of the `conditional evolution' of the Kondo 
singlet under
projective measurement on the detector. Our study reveals that the state of
the composite system is disentangled upon this measurement. The Kondo
singlet evolves into a particular state with a fixed number of electrons
in the quantum dot. Its relaxation time is shown to be sensitive only
to the QD-charge dependence of the transmission probability in the detector,
which implies that the phase information is erased in this conditional
evolution process.
We discuss implications of our observations in view of the possible
experimental realization.

\end{abstract}

\maketitle

\section{Introduction}
Quantum interference and its suppression caused by interactions with external
degrees of freedom have been central subjects of mesoscopic physics for
more than a decade~\cite{dittrich98}. These subjects deal with the 
transition from quantum to classical phenomena in mesoscopic scales. 
In particular, ``which-path" (WP) detection
in mesoscopic quantum interferometers 
provides an ideal playground for studying the
complementarity (which is often identified with `wave-particle duality')
in quantum theory. Experiments on the
controlled dephasing have been performed in mesoscopic structures based
on quantum dots (QD)~\cite{buks98,sprinzak00,kalish04}. 
A prototype experimental setup for this
kind of study~\cite{buks98} is as follows. 
Coherent transmission of
electrons is monitored by using an Aharonov-Bohm (AB)
interferometer with a QD inserted in one of the interferometer's 
arms~\cite{yacoby95,schuster97}. A mesoscopic detector 
is electrostatically coupled to the QD. Because of electrostatic interactions, 
the electron state in the detector depends on the charge state of the QD, 
which results
in a quantum correlation (i.e., ``entanglement") between the QD and the 
detector. 
The AB oscillation of the conductance through the interferometer is 
suppressed because of the WP information transferred to the detector.
This ``measurement-induced dephasing" is controlled through the
voltage applied across the mesoscopic detector. The
controlled dephasing experiments were carried out also without an
AB interferometer~\cite{sprinzak00,kalish04}, 
because it is possible to study the coherence 
by the resonant transmission through a QD. Various theoretical approaches
were used to study this 
problem~\cite{aleiner97,levinson97,gurvitz97,stodolsky99,buttiker00}.

The controlled dephasing experiment was also
performed in the Kondo limit of the QD~\cite{kalish04}. A Kondo
singlet is formed between the localized spin in a QD and
electrons in the leads~\cite{hewson93}, which gives rise to
enhanced transport through the
QD~\cite{gordon98,cronen98,schmid98,simmel99,ji00,wiel00,kouwenhoven01}.
It was shown that a nearby quantum point contact (QPC) capacitively coupled to 
the QD plays a
role of a ``potential detector" and suppresses
the Kondo resonance~\cite{kalish04}.
However, characteristics of the
measured suppression were very different from the theoretical
prediction of Ref.~\cite{silva03}. The most significant deviation from
the theory is that the measured suppression strength is much
larger (about 30 times) than expected. Dependence on the
transmission probability ($T$) and on the bias voltage ($V$) across
the QPC were also inconsistent with the theoretical expectation.
The analysis of the experiment~\cite{kalish04} was based on a
theory~\cite{silva03} of dephasing of the Kondo resonance as a result
of path detection by the QPC through the change of the
transmission {\em probability}, $\Delta T$. It was pointed out
that this kind of treatment does not fully take into account the 
WP information acquired in the detector~\cite{kang05}. 
It is because scattering of electrons at
the QPC is a quantum mechanical phenomenon with complex
transmission and reflection amplitudes. Therefore, in general, {\em
phase-sensitive} information should also be taken into
account~\cite{sprinzak00,stodolsky99,kang05,hacken01}.

In this paper, first we present a theory of the entanglement of
the Kondo singlet with a mesoscopic detector (Section 2). 
The formulation is based on
the variational ground state of the Kondo singlet~\cite{hewson93,gunnar83}
correlated with the charge-dependent detector state. We, then, report
on our investigations of
dephasing in the Kondo state controlled by charge detection (Section 3).
Discussions in Section 2 and 3 are mainly extension of the study in 
Ref.~\cite{kang05}. In addition,
a refined model for the detector is introduced which explains the importance
of the phase-sensitive WP information. In Section 4, we introduce the concept
of the ``conditional evolution" of the Kondo singlet under projective
measurement on the detector. We show that the phase-sensitive 
information is erased and the Kondo state suffers relaxation in
a way that depends on the charge sensitivity of the detector current. 
Conclusion is given in Section 5. Also, the relation between the scattering
matrix and the parallel shift of the one-dimensional potential is derived
in the Appendix. 

\section{The entanglement of the Kondo singlet with a charge detector}
\begin{figure}
\centering%
\includegraphics{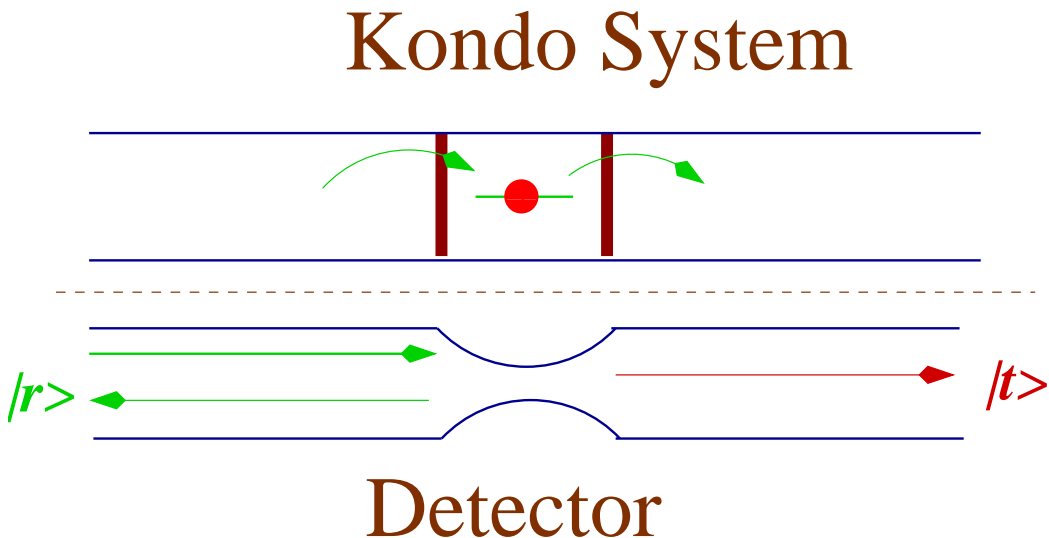}
\caption{Schematic figure of the Kondo system interacting with a detector. The
Kondo system is composed of a quantum dot connected to two electrodes. For 
the detector we consider a two-terminal mesoscopic conductor with a single 
transmission channel.
 }
\end{figure}
The model system we investigate is schematically drawn in Figure~1.
First, to describe the Kondo singlet of the QD, we adopt the variational
ground state for the impurity Anderson
model~\cite{hewson93,gunnar83}. This variational ground state captures 
the essential Kondo physics in a simple but effective way.
Furthermore, this approach can be easily applied for describing the
entanglement of the Kondo singlet with the detector. The Hamiltonian 
for the QD + two electrodes + tunneling is given by
\begin{equation}
H = H_L + H_R + H_D + H_T \;.
\end{equation}
The left (L) and the right (R) leads are described by the noninteracting
Fermi sea as
\label{eq:hamil}
\begin{equation}
H_\alpha = \sum_{k\sigma}\varepsilon_{\alpha k}
c^\dagger_{\alpha k\sigma} c_{\alpha k\sigma} \quad
(\alpha = L, R) \,,
\end{equation}
where $c_{\alpha k\sigma}$ $(c_{\alpha k\sigma}^\dagger)$ is an
annihilation (creation) operator of an electron with energy
$\varepsilon_{\alpha k}$, momentum $k$, and spin $\sigma$ on the
lead $\alpha$.
The interacting QD is represented by $H_D$ given as
\begin{equation}
H_D = \sum_\sigma \varepsilon_d d_\sigma^\dagger d_\sigma
+ U n_\uparrow n_\downarrow  \,,
\end{equation}
where $d_\sigma$ and $d_\sigma^\dag$ are the QD electron annihilation and 
creation operators, respectively, and $n_\sigma=d_\sigma^\dag d_\sigma$. 
The parameters, $\varepsilon_d$ and $U$, stand for the
energy of the localized level and the on-site Coulomb interaction,
respectively.
The tunneling Hamiltonian $H_T$ has the form
\begin{equation}
\label{kondo-phase:1c}
H_T = \sum_{\alpha=L,R}\sum_{k\sigma}
\left(V_\alpha d_\sigma^\dag c_{\alpha k\sigma} + h.c.\right) \;,
  \end{equation}
where $V_\alpha$ is responsible for the tunneling between the QD and
  the lead $\alpha$.

In the absence of interaction between the QD and the detector, the variational
ground state for the Hamiltonian $H$ ($U\rightarrow\infty$ limit) is written
as~\cite{hewson93,gunnar83}
\begin{equation}
 |\Psi_G\rangle = A|0\rangle + B|1\rangle,
\end{equation}
where $|0\rangle$ denotes the Fermi sea for the leads with an empty QD state, 
and
\begin{equation}
   |1\rangle \equiv  \frac{1}{\sqrt{2}} \sum_{\alpha\sigma,k<k_F}
   v_{\alpha k} d_\sigma^\dagger c_{\alpha k\sigma} |0\rangle \,.
\end{equation}
Here $A=\sqrt{1-{n_d}}$ and $B=\sqrt{n_d}$, with ${n_d}$
being the average occupation
number of the QD level and 
\begin{equation}
v_{\alpha k} = \sqrt{\frac{2n_d}{1-n_d}} 
  \frac{V_\alpha}{E_G-\varepsilon_d +\varepsilon_{\alpha k}},
\end{equation}
where $E_G$ denotes the ground state energy determined by the
equation
\begin{equation}
 E_G = 2\sum_{\alpha,k<k_F}
       \frac{ V_\alpha^2 }{ E_G-\varepsilon_d+\varepsilon_{\alpha k} } \;.
\end{equation}
The Kondo temperature ($T_K$), the characteristic energy scale of the
system, is given as a difference between the QD level
($\varepsilon_d$) and the ground state energy ($E_G$):
$T_K=\varepsilon_d-E_G$.
%

In fact, the states $|0\rangle$ and $|1\rangle$ have different
occupation numbers for the QD: $n_d=0$ and $n_d=1$, respectively.
A detector (usually a mesoscopic conductor) near to the QD is able to 
detect the charge
state, since the potential of the detector depends on the charge state
of the QD. So the transmission and reflection amplitudes of
the detector also depend on $n_d$. This correlation can be described by an
entangled state for the composite system as
\begin{equation}
 |\Psi_{tot}\rangle = A |0\rangle\otimes|\chi_0\rangle 
   + B |1\rangle\otimes|\chi_1\rangle ,
\label{eq:Psi_tot}
\end{equation}
where $|\chi_i\rangle$ ($i=0$ or $i=1$) denotes the detector state
when the Kondo system is in the state $|i\rangle$. Here, 
a two-terminal single-channel conductor is considered as the detector. 
Then, an injected electron from the left electrode of the detector can
be described by the state 
\begin{equation}
 |\chi_i\rangle = r_i|r\rangle + t_i|t\rangle, 
\end{equation}
where $r_i$ and $t_i$ are the $i$-dependent reflection and transmission 
coefficients, respectively. The state $|r\rangle$ ($|t\rangle$) corresponds 
to the state of reflection (transmission) for an injected electron.
It is important to note that the scattering coefficients are complex
numbers that can be expressed as
\begin{eqnarray}
 r_i &=& |r_i|\exp{(i\phi_{r_i})},\\
 t_i &=& |t_i|\exp{(i\phi_{t_i})}, 
\end{eqnarray}
and satisfy the unitarity relation
$|r_i|^2+|t_i|^2=1$.
\section{Current- and phase-sensitive dephasing}
\subsection{The reduced density matrix for the Kondo system}
Dephasing of the Kondo singlet takes place when an observer (detector)
monitors the
Kondo system (which is a part of the composite system). This can be 
described in terms of the reduced
density matrix approach. Before interaction of the two subsystems, the
density matrix of the Kondo singlet is given as 
$\rho_0=|\Psi_G\rangle\langle\Psi_G|$. Upon a single scattering event
with the detector, the reduced density matrix  
$\rho$ of the Kondo system is given as 
\begin{equation}
 \rho = \mathsf{Tr}_\mathsf{det} \left(|\Psi_{tot}\rangle\langle\Psi_{tot}| 
   \right),
\end{equation}
where $\mathsf{Tr}_\mathsf{det}(\cdots)$ denotes a trace over the detector
degree of freedom.

It is found that the diagonal elements of $\rho$ do not change
upon scattering at the detector. On the other hand, the off-diagonal 
elements are modified by
\begin{equation}
 \rho_{01} = \lambda\rho_{01}^0  ,
\end{equation}
where
\begin{equation}
 \lambda = \langle\chi_1|\chi_0\rangle 
   = r_0r_1^* + t_0t_1^* . \label{eq:lambda}
\end{equation}
This quantity represents the information of the charge state in the QD 
transferred to the detector. That is, $\lambda=0$ implies that the two
states are orthogonal. Thus, a complete charge-state information is 
transferred to the detector. For $|\lambda|=1$, the two charge states 
are identical which means that the detector does not obtain any information
on the QD state.
Dephasing of the Kondo singlet takes place for $|\lambda|<1$. It is 
obvious from
Equation~(\ref{eq:lambda}) that the dephasing is associated
not only with the current sensitivity but also with the phase sensitivity
of the scattering coefficients of the detector. 
\subsection{The time evolution of the density matrix 
 in the weak measurement limit}
We consider the weak continuous measurement limit where the scattering 
through the detector takes place on a time scale much shorter than the 
relevant time scales in the Kondo 
singlet.  In our case, $\Delta t \ll t_d$, where $\Delta t=h/2eV$ denotes the
average time between two successive scattering events with $V$
being the voltage applied across the detector. 
The parameter, $t_d$, is the
dephasing time of the Kondo singlet. 
This assumption allows us to use the Markov approximation that neglects the
memory effect in the detector. Then after the scattering of $n$ electrons
through the detector the off-diagonal component of the
density matrix is given as
\begin{equation}
 \rho_{01}(t) = \lambda^n \rho_{01}(0) \,,
\end{equation}
where $t=n\Delta t$. Note that the time evolution of the density matrix
here is written in the Heisenberg picture in order to eliminate the
less important dynamical phase factor. 
This equation can be rewritten in the form
\begin{equation}
 \rho_{01}(t) = e^{ (i\Delta\epsilon-\Gamma_d) t }
 \rho_{01}(0) \,,
\end{equation}
where the two parameters, $\Delta\epsilon$ and $\Gamma_d$, 
represent the phase shift and the dephasing
rate, respectively, caused by the detection processes.
One can find that
\begin{eqnarray}
 \Delta\epsilon  &=& \frac{2eV}{h}\arg{\lambda} \,, \\
 \Gamma_d &=& \frac{1}{t_d} = -\frac{2eV}{h} \log|\lambda| \;. \label{eq:1/td}
\end{eqnarray}
On the other hand, the diagonal terms are independent of time.
This implies that no relaxation takes place in the Kondo singlet.

In the weak measurement limit, $\lambda\sim1$, $\Gamma_d$ and $\Delta\epsilon$
can be expressed in terms of the changes in the magnitude and phase
of the scattering amplitudes of the detector as
\begin{equation}
 \Gamma_d = \Gamma_T + \Gamma_\phi
\label{eq:Gamma_d}
\end{equation}
where
\begin{eqnarray}
 \Gamma_T &=& \frac{eV}{h} \frac{(\Delta T)^2}{4T_0(1-T_0)} \;, 
  \label{eq:Gamma_T}\\
 \Gamma_\phi &=& \frac{eV}{h} T_0(1-T_0) (\Delta\phi)^2 \;,
  \label{eq:Gamma_phi}\\
\end{eqnarray}
and
\begin{equation}
 \Delta\epsilon = \frac{eV}{\pi}(1-T_0)\Delta\phi_r
     + \frac{eV}{\pi} T_0\Delta\phi_t \;.
\end{equation}
Here, $T_0=|t_0|^2$ ($T_1=|t_1|^2$) is the transmission probability in the
absence (presence) of an extra electron in the QD. 
$\Delta T\equiv |t_0|^2-|t_1|^2=|r_1|^2-|r_0|^2$
represents the change in the transmission probability. 
The phase shift $\Delta\phi$ is given by 
$\Delta\phi=\Delta\phi_t-\Delta\phi_r$, where $\Delta\phi_t$
($\Delta\phi_r$) is the change in the transmission
(reflection) amplitude resulting from the different charge states:
$\Delta\phi_t=\phi_{t0}-\phi_{t1}$,
$\Delta\phi_r=\phi_{r0}-\phi_{r1}$.

%
\subsection{Dephasing and the Kondo-assisted transport}
\begin{figure}
\centering%
\includegraphics{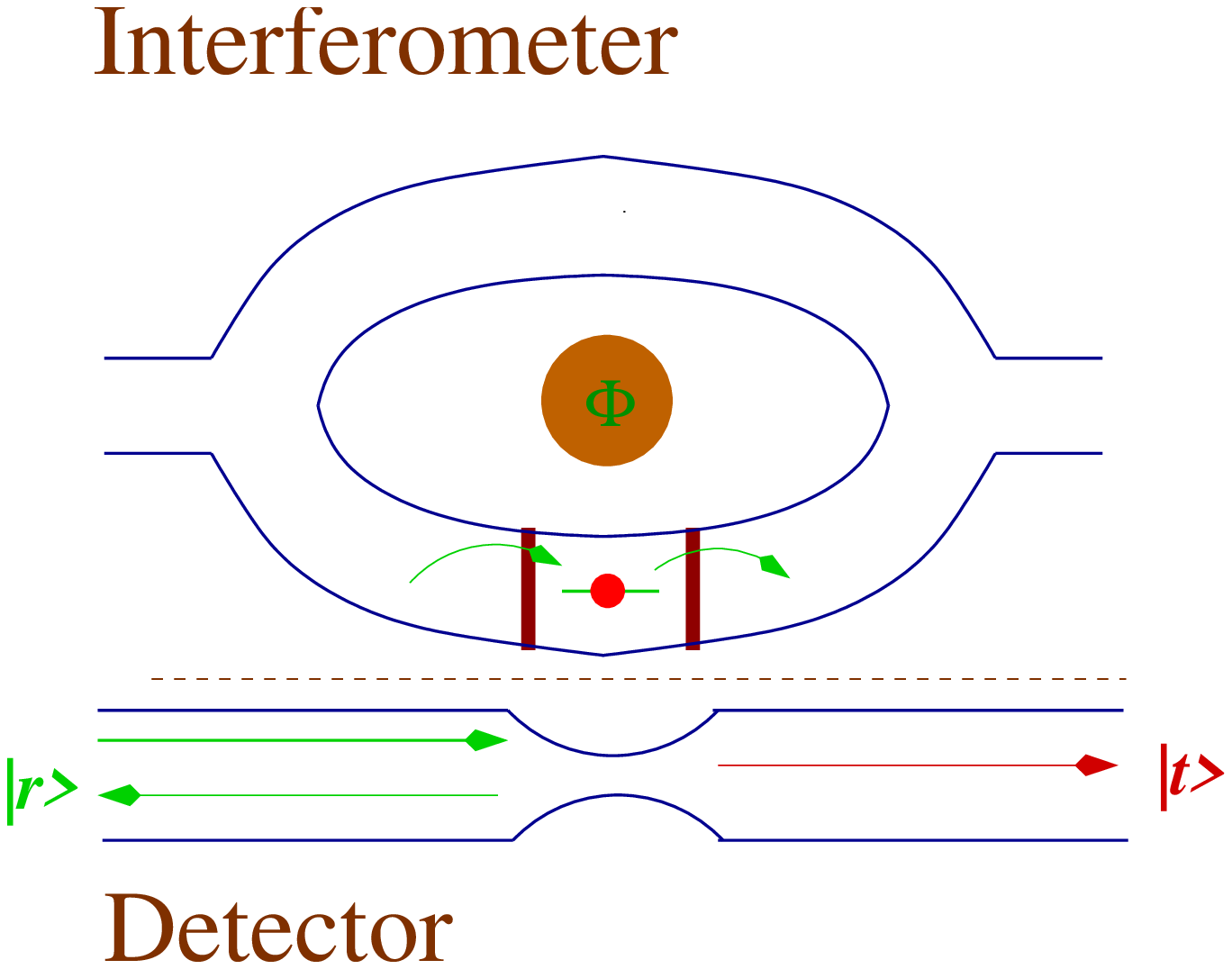}
\caption{Schematic figure of an Aharonov-Bohm interferometer with a detector.
A quantum dot in the Kondo regime is inserted in the interferometer. The 
electrons in the quantum dot interact with the detector electrons.
 }
\end{figure}

The effect of dephasing can be investigated through electron transport in
the Kondo system. The best way for studying the dephasing would be to
compose a two-path AB interferometer with a Kondo-correlated QD inserted
in one arm of the interferometer~\cite{ji00} (Figure~2). In this case, 
the total transmission probability ($T_{AB}$) through the interferometer 
is given as
\begin{equation}
 T_{AB} = |t_{ref}+t_{QD} e^{i\theta}|^2 = |t_{ref}|^2 + |t_{QD}|^2 
        + 2|t_{ref}||t_{QD}|\cos{\theta} ,
\label{eq:TAB}
\end{equation}
where $t_{ref}$ and $t_{QD}$ stand for the transmission amplitudes through
the reference arm and through the QD, respectively. The relative phase
shift $\theta$ is controlled by the external AB flux $\Phi$ as
\begin{displaymath}
 \theta = \frac{2\pi e\Phi}{hc} + const.
\end{displaymath}
The magnitude of the AB oscillation in Equation~(\ref{eq:TAB}), 
denoted by ${\cal V}_{AB}$, is given as
\begin{equation}
 {\cal V}_{AB} = 2|t_{ref}||t_{QD}| .
\end{equation}
The measurement-induced dephasing
is expected to reduce $|t_{QD}|$. For a more quantitative study, we use the
following relationship between $t_{QD}$ and Green's function for the QD, 
$G_d(\omega)$, at the Fermi energy ($\omega=0$) as~\cite{langreth66} 
\begin{equation}
 t_{QD} = -2i\sqrt{\Gamma_L\Gamma_R} G_d(0),
\end{equation} 
where $\Gamma_L$ ($\Gamma_R$) is the tunneling rate of an electron
between the QD and the left (right) electrode.
Green's function for the mixed state described by the
reduced density matrix $\rho$ (see the previous section) is defined by
\begin{equation}
 G_d(\omega) = -i\int_0^\infty dt\,e^{i\omega t} \mathsf{Tr}\left(
  \rho(t) [d_\sigma(t),d_\sigma^\dagger]_+
  \right) \;,
\end{equation}
where $[\cdots,\cdots]_+$ denotes the anti-commutator.
Green's function can be evaluated in a similar way to
the one in Ref.~\cite{hewson93,gunnar83}.
We need to use the equations of motion for various
Green's functions and truncate higher order terms of $1/N_s$ with $N_s$
being the spin degeneracy. Neglecting the incoherent background and the energy
shift $\Delta\epsilon$, we obtain the following expression 
in the Kondo limit ($n_d\sim1$)
\begin{equation}
 G_d(\omega) \simeq \frac{(1-{n_d})}{\omega-T_K + i\Gamma_d}.
  \label{eq:green}
\end{equation}
Therefore we find that the magnitude of the AB oscillation is reduced
by the dephasing when the voltage $V$ is applied in the detector by the
factor
\begin{equation}
 \frac{{\cal V}_{AB}(V)}{ {\cal V}_{AB}(V=0) }
   = \sqrt{\frac{T_K^2}{T_K^2+\Gamma_d^2} }.
\end{equation}
Note that the $V$-dependence of
${\cal V}_{AB}$ comes through the relation~(\ref{eq:1/td}).
 
Alternatively, one can study the dephasing of the Kondo singlet through
direct transport through the QD without interferometry (Figure~1). 
The phase coherence
of the Kondo state appears in the resonant transport through the 
double-barriers which is an electronic analogue of the Fabry-Perot
interferometer~\cite{optics84}. The experiment
carried out in Ref.~\cite{kalish04} used this geometry. In this case,
the conductance is proportional to $|t_{QD}|^2$, leading to the suppression
of the conductance at a finite detector bias $V$, by the factor 
\begin{equation}
 g(V) = \frac{T_K^2}{T_K^2+\Gamma_d^2}. 
\end{equation}  

\subsection{The dephasing rate and symmetric vs. asymmetric charge responses}
From our discussion, it is obvious that the Kondo
resonance is reduced by the charge detection of the QD through the coherent
scattering at the detector. This coherent scattering is described by
the complex transmission and reflection coefficients. In the weak
measurement limit, the dephasing rate is given by the
phase-sensitive ($\Delta\phi$) as well as the current-sensitive
($\Delta T$) detection. 

The phase-sensitive contribution to dephasing $\Gamma_\phi$ was not taken 
into account 
in the experimental report of Ref.~\cite{kalish04}.
The much stronger dephasing rate than expected in the theory in 
Ref.~\cite{silva03} 
which takes only $\Gamma_T$ into account suggests a large contribution from the
phase-sensitive dephasing, i.e., $\Gamma_\phi\gg\Gamma_T$. 
One of the authors (K.~K.) has pointed out previously~\cite{kang05} that 
phase-sensitive dephasing
might be dominant in a generic situation if the asymmetry in the charge
sensitivity of the detector potential is taken into account.
Here, we provide a refined version of the detector model for demonstrating
this behavior.
First, $\Delta\phi=0$ if the detector potential and its variation resulting
from an
extra QD electron have inversion
symmetries~\cite{korotkov01,pilgram02} and thus the phase-sensitive
contribution vanishes~\cite{note}.
However, in reality, there is no reason to believe that the
response of the detector potential to the QD charge should be
symmetric. We take into account the asymmetric as well as the symmetric 
variation of the detector potential.
The potential profile $V_i(x)$ depends on the charge state of the QD $i\in0,1$.
We use a one-dimensional inverse harmonic potential for the detector as
(See Figure~3)
\begin{eqnarray}
 V_0(x) &=& V_0 - \frac{1}{2}m\omega_x^2x^2 ,\\
 V_1(x) &=& V_0 +\delta V_0 - \frac{1}{2}m\omega_x^2(x-\delta x)^2.
\label{eq:qpc-potential}
\end{eqnarray}
The parameter $\delta V_0$ corresponds to the symmetric component of the 
charge sensitivity.
Its asymmetry is accounted for by the parallel shift $\delta x$
of the potential profile.
The transmission probability $T_i$ ($i\in 0,1$) can be exactly calculated
in this model~\cite{fertig87}. We find that
\begin{eqnarray}
 T_0 &=& \frac{1}{1+\exp{(-2\pi\varepsilon_0)}} ,\\
 T_1 &=& T_0-\Delta T = \frac{1}{1+\exp{(-2\pi(\varepsilon_0-\delta v_0))}} ,
\end{eqnarray}
 where the dimensionless variables $\varepsilon_0$ and $\delta v_0$ are
 defined by
 \begin{equation}
  \varepsilon_0 = \frac{E-V_0}{\hbar\omega_x} ,\;\;\;
  \delta v_0 = \frac{\delta V_0}{\hbar\omega_x} .
  \end{equation}
In the weak measurement limit ($\delta v_0\ll1$), we find that
\begin{equation}
 \Delta T \simeq 2\pi T_0(1-T_0) \delta v_0 .
\end{equation}
$\delta V_0$, the symmetric component of the potential response,
does not contribute to the phase-sensitive 
dephasing~\cite{korotkov01,pilgram02,note}.
Then the phase sensitivity of the detector is purely given by the
parallel shift $\delta x$ of the potential as (see Appendix)
\begin{equation}
 \Delta\phi = 2 k_F\delta x ,
 \label{eq:Delta-phi}
\end{equation}
where $k_F$ denotes the Fermi wave vector.
Therefore, for the potential profile used in Equation~(\ref{eq:qpc-potential}),
the dephasing rates are given by
 \begin{eqnarray}
 \Gamma_T &=& \pi^2\frac{eV}{h} T_0(1-T_0)(\delta v_0)^2, \\
\Gamma_\phi &=& 4\frac{eV}{h} T_0(1-T_0)(k_F\delta x)^2.
\end{eqnarray}

\begin{figure}
\centering%
\includegraphics{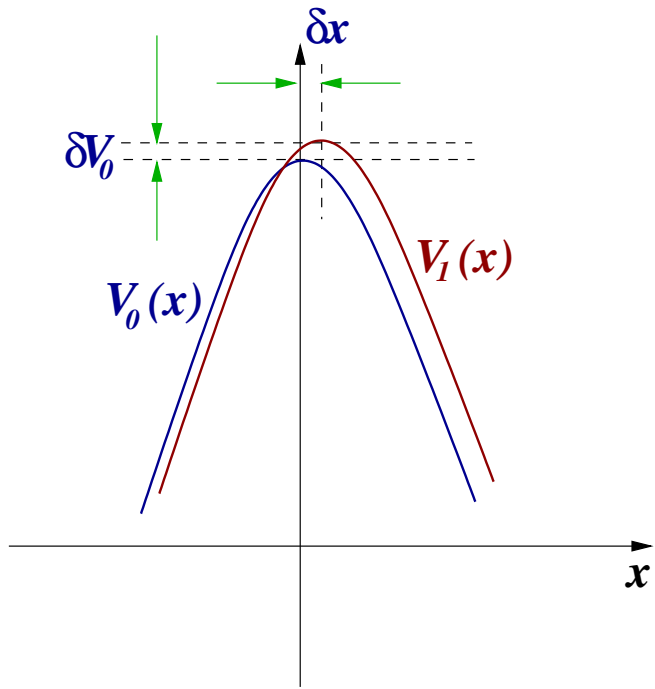}
\caption{A model for the detector potentials $V_i(x)$ depending on the
charge state $i$ ($\in0,1$) of the quantum dot. The symmetric and the 
asymmetric responses to an extra electron of the quantum dot are taken into
account via $\delta V_0$ and $\delta x$, respectively.
 }
\end{figure}

In the case where the asymmetric response of the potential is comparable to 
the symmetric response, the following relation will be satisfied:
\begin{displaymath}
 \frac{\delta V_0}{\hbar\omega_x} \sim \left(\frac{\delta x}{x_0}  \right)^2 ,
\end{displaymath}
where
$x_0\equiv \sqrt{\hbar/m\omega_x}$ characterizes the length
scale of the detector potential. This relation implies that the changes of 
the energy 
scales resulting from the symmetric and the asymmetric responses are comparable.
With this condition, we get
\begin{equation}
 \Gamma_\phi = 4\frac{eV}{h} T_0(1-T_0)(k_F x_0)^2 \delta v_0. 
\end{equation}
In the experiment of Ref.~\cite{kalish04}, the Fermi wavelength ($\lambda_F=
2\pi/ k_F$) and the length scale of the detector potential are about
$\lambda_F\sim 44$nm and $x_0\sim {\cal O}(20$nm), respectively. Therefore, 
$k_F\delta x\sim k_F x_0\sqrt{\delta v_0} \sim 3 \sqrt{\delta v_0}$  
and we find that
\begin{equation}
 \frac{\Gamma_\phi}{\Gamma_T} \sim (\delta v_0)^{-1}.
\end{equation}
Because $\delta v_0 \ll1$ in our description,
the condition $\Gamma_\phi\gg\Gamma_T$ can be achieved if the
asymmetric response in the detector potential is not negligible.
This conclusion can also be understood as follows. The 
transmission probability is affected
across the region $|x|\lesssim x_0$, while the phase is affected
through a relatively wide region, so that the
phase-sensitive detection is more effective. This leads to a large
contribution of phase-sensitive dephasing 
and can be a natural explanation for the anomalously large dephasing rate
observed in Ref.~\cite{kalish04}.

It should be noted that our discussion on the large phase-sensitive dephasing
is not restricted to the Kondo limit. 
However, in the Kondo limit, electrons in the Kondo cloud may interact with
the electron in the detector. We expect that this interaction rarely
affects the transmission probability. But it contributes to 
the phase-sensitive
dephasing. Interactions with the Kondo cloud
is expected to increase the asymmetric response of the detector potential.
This argument could explain why the phase-sensitive dephasing
is more pronounced in the Kondo limit than in the Coulomb blockade limit.

We also briefly remark on the $T_0$-dependence of
$\Gamma_d$. For the simple model of the detector considered here, 
$\Gamma_d$ is expected to be
proportional to the partition noise ($\propto T_0(1-T_0)$) of the
ideal single-channel detector. However, the experimental $\Gamma_d$-$T_0$
curve shows a double peak behavior~\cite{kalish04} in contrast to
the theoretical model. This qualitative discrepancy might be
related to the so called ``0.7 anomaly"~\cite{0.7anomaly} where
the shot noise is also suppressed~\cite{roche04} or to the charge
screening effect~\cite{buttiker00,buttiker03}. This issue
requires more careful experimental and theoretical analysis on the
correlation between the dephasing of the Kondo state and the shot noise
of the detector.

\section{The conditional evolution of the Kondo state}
In this section, we investigate the time evolution of the Kondo singlet 
{\em conditioned on the observation of a particular measurement} on the
detector. The conditional dynamics 
of a state are obtained by an operation on a part of the system that 
corresponds
to a specific classical outcome of measurement and renormalizing the 
reduced wave function so that it has a total probability of one (see 
e.g. Ref.~\cite{nielson00}). In the case of a two-terminal mesoscopic detector,
there are two possible outcomes of measurement on the detector, that is, 
transmission
and reflection, for each of the injected electrons~\cite{averin06}. These
measurement processes are described by the operators $\hat{M}_t$
and $\hat{M}_r$ defined as
\begin{equation}
  \hat{M}_t =
   \frac{|t\rangle\langle t|}{ 
      \sqrt{\langle\Psi_{tot}|t\rangle\langle t|\Psi_{tot}\rangle } }, \;\;\;
  \hat{M}_r = 
   \frac{|r\rangle\langle r|}{ 
      \sqrt{\langle\Psi_{tot}|r\rangle\langle r|\Psi_{tot}\rangle } }.
\end{equation}
Upon a measurement $\hat{M}_t$ 
the state $|\Psi_{tot}\rangle$ of Equation~(\ref{eq:Psi_tot})
is reduced to $|\Psi^t\rangle$ as
\begin{equation}
 |\Psi^t\rangle = \hat{M}_t|\Psi_{tot}\rangle 
  = \left( A'|0\rangle + B'|1\rangle \right) \otimes |t\rangle ,
\label{eq:Psi_t}
\end{equation}
where
\begin{equation}
 A' = \frac{At_0}{ \sqrt{|A|^2T_0 + |B|^2T_1} }, \;\;\;
 B' = \frac{Bt_1}{ \sqrt{|A|^2T_0 + |B|^2T_1} }.
\end{equation}
Unlike  the state $|\Psi_{tot}\rangle$, the state $|\Psi^t\rangle$ of 
Equation~(\ref{eq:Psi_t}) is not entangled but expressed as a product
state of the Kondo system and the detector. That is,
under the projective measurement of the detector, the two subsystems
are disentangled. This is because the measurement $\hat{M}_t$ selects
one of the two possible outcomes of the detector and the detector electron
is collapsed onto the state $|t\rangle$. This means that we can describe
the Kondo state under measurement $\hat{M}_t$ by
a pure state $ A'|0\rangle + B'|1\rangle $. Therefore, under a continuous
weak measurement, we can write the conditional evolution of the
Kondo singlet as
\begin{equation}
 |\Psi_G^t(t)\rangle = A(t)|0\rangle + B(t)|1\rangle ,
 \label{eq:Psi_G}
\end{equation}  
where the time evolution of the amplitudes $A(t)$ and $B(t)$ satisfy the
relations (upon a mean time interval $\Delta t_T\equiv h/2eVT_0$ between 
successive transmissions of electrons in the detector)
\begin{eqnarray}
 A(t+\Delta t_T) &=& \frac{t_0}{ \sqrt{|A(t)|^2T_0 + |B(t)|^2T_1} } A(t), \\
 B(t+\Delta t_T) &=& \frac{t_1}{ \sqrt{|A(t)|^2T_0 + |B(t)|^2T_1} } B(t),
\end{eqnarray}
or simply one can find that
\begin{equation}
 \frac{B(t+\Delta t_T)}{A(t+\Delta t_T)} = \frac{t_1}{t_0} \frac{B(t)}{A(t)}.
\end{equation}
This relation gives the time evolution of the ratio between the two
coefficients as
\begin{equation}
 \frac{B(t)}{A(t)} = \exp{[(-\Gamma_{rel}/2+i\eta)t]} ,
\end{equation}
where 
\begin{eqnarray}
 \Gamma_{rel} &=& -\frac{2eVT_0}{h}\log{(1-\frac{\Delta T}{T_0})}, \\
 \eta &=& \frac{2eVT_0}{h} \Delta\phi_t.
\end{eqnarray}
Therefore the relative probability of the two states 
$|B(t)|^2/|A(t)|^2$ goes to zero at $t\rightarrow\infty$ as
\begin{equation}
 \frac{|B(t)|^2}{|A(t)|^2} = \exp{(-\Gamma_{rel}t)} 
 \label{eq:BoverA}
\end{equation} 
This result implies that the Kondo state evolves into the state $|0\rangle$
at $t\rightarrow\infty$ with its relaxation rate $\Gamma_{rel}$.
In the weak measurement limit $\Delta T/T_0$ must be much smaller than
unity and the relaxation rate is simplified as
\begin{equation}
 \Gamma_{rel} \simeq \frac{2eV}{h} \Delta T .
 \label{eq:Gamma_rel}
\end{equation}

Several interesting observations can be made from 
Equations~(\ref{eq:BoverA},\ref{eq:Gamma_rel}).
First, the relaxation rate is sensitive only to the change of the transmission
probability. It is independent of the phase shift.
This is in strong contrast with the dephasing rate $\Gamma_d$ of 
Equations~(\ref{eq:Gamma_d}-\ref{eq:Gamma_phi}) where the 
charge-state information is contained both in
the change of the transmission probability and the phase shift.
In other words, the measurement $\hat{M}_t$ on the state $|\Psi_{tot}\rangle$ 
washes out part of the charge-state information encoded in the phase shift.
Indeed, for a detector sensitive only to the scattering 
phases (that is, for $\Delta T=0$), the Kondo singlet of 
Equation~(\ref{eq:Psi_G})
remains unchanged (aside from the phase factor $e^{i\eta t}$). In this
case the phase coherence of the Kondo singlet is fully 
preserved. In fact, this corresponds to the quantum erasure of the 
charge-state information by a particular measurement ($\hat{M}_t$ in our case)
on the 
detector~\cite{kang06}. The time evolution of the Kondo singlet, conditioned
on the measurement $\hat{M}_t$, does not show any relaxation if the detector
current is not sensitive to the charge state of the QD. 

In an experiment, this conditional evolution and relaxation of the Kondo 
singlet
can be investigated by correlating the transport of the electron through the QD 
and detection of electron at the output lead of the detector. 
For instance, let us consider an AB interferometer with a QD embedded in
one of its arms and a detector nearby the QD as discussed in Section 3
(Figure~2). The Kondo-resonant transport under the measurement 
$\hat{M}_t$ can be studied through the  
zero-frequency cross-correlation measurement  
between the two output leads, one from the interferometer and the other
from the detector (See e.g.,~\cite{kang06}).
In this case, the interference in the cross-correlation will be reduced
in proportion to the relaxation rate $\Gamma_{rel}$. As discussed above,
the suppression of the interference is related to the charge sensitivity of
the detector in transmission probability. The phase-sensitive information
would not affect the visibility in the joint detection of the electrons
at the two output electrodes.

We also point out that the cross-correlation measurement is able to resolve
the anomaly observed in a controlled dephasing experiment of the
Kondo-correlated QD~\cite{kalish04}. The experimental results show
unusually larger than expected dephasing rate with theory~\cite{silva03}
based on `current-sensitive' dephasing which is equivalent to
$\Gamma_T$ in Equation~(\ref{eq:Gamma_T}). As we have shown in Section 3.4,
the phase-sensitive contribution of dephasing
can be dominant (that is, $\Gamma_\phi\gg\Gamma_T$) by taking into account 
asymmetry in the potential response. The relaxation rate $\Gamma_{rel}$ 
does not
contain the phase shift of the scattering coefficients. Therefore, by
measuring the conditional count on the Kondo-resonant transmission,
one can extract the value $\Delta T$. Therefore, by combining the 
cross-correlation and the usual current measurement, we can get
the two different contributions of dephasing $\Gamma_T$ and
$\Gamma_\phi$. This would be a direct way to confirm the theoretical
prediction on the importance of the contribution of asymmetry in the detector.
 
Aside from the phase-sensitive contribution to dephasing, $\Gamma_\phi$,
it is interesting to note that $\Gamma_{rel}\ne\Gamma_T$.
Let us consider a system with perfect inversion symmetry (thus $\Gamma_\phi=0$)
in the detector. For the weak continuous measurement considered in our
study, we can find that $\Gamma_{rel}\gg\Gamma_T$. In other words,
the interference in the Kondo-assisted transmission 
under the measurement $\hat{M}_t$
is reduced much faster than in the case without the measurement. 
This can be regarded as an interesting manifestation of the nonlocality
of quantum theory.

So far in this section we have discussed conditional evolution under the 
measurement $\hat{M}_t$.
The same kind of investigation can be done for the measurement $\hat{M}_r$. 
One can find that the Kondo singlet under this measurement (denoted by 
$|\Psi_G^r(t)\rangle$) evolves into the 
state $|1\rangle$ as
\begin{equation}
 |\Psi_G^r(t)\rangle = \bar{A}(t)|0\rangle + \bar{B}(t)|1\rangle,
\end{equation}
where the two coefficients $\bar{A}(t)$ and $\bar{B}(t)$ satisfy the relation
\begin{equation}
 \frac{|\bar{B}(t)|^2}{|\bar{A}(t)|^2} = \exp{(\Gamma_{rel}t)} 
\end{equation}
with its relaxation rate $\Gamma_{rel}$ 
being equivalent to the one obtained in Equation~(\ref{eq:Gamma_rel}): 
\begin{displaymath}
 \Gamma_{rel} \simeq \frac{2eV}{h} (|r_1|^2-|r_0|^2) = \frac{2eV}{h} \Delta T. 
\end{displaymath}

An important point from our observation is that the present charge
detection process should not be considered as an irreversible phase 
randomization which may be present because of some uncontrollable degrees
of freedom. As described above, cross-correlation measurements can be
used to recover the interference and therefore clarify that it cannot be
attributed to 
irreversible phase randomization. Another kind of interferometer+detector 
setup has also been investigated that is able to confirm
this point of view~\cite{khym06}. 
\section{Conclusion}
We have described the Kondo singlet in a quantum dot 
entangled with a mesoscopic charge detector. 
Without any `measurement' on the detector, 
the `coherence' of the Kondo singlet
is reduced. The dephasing rate is sensitive to the charge-state information
encoded both in the magnitude and in the phase of the scattering coefficients
of the detector. A detector model is introduced to account for the two
different contributions of dephasing and to provide a possible solution
to a recent experimental puzzle~\cite{kalish04}. 
In the case that projective measurements are performed on the detector 
electrodes, the Kondo singlet is disentangled from the detector state.
In this case, the Kondo
singlet evolves into a particular state with a fixed number of electrons
in the quantum dot. Its relaxation rate is shown to be sensitive only
to the QD-charge dependence of the transmission probability in the detector.
This implies that the phase information is erased in the conditional
evolution process. This kind of relaxation can be investigated by a 
cross-correlation measurement on the two output electrodes, one from the Kondo
system and the other from the detector.

\section*{Appendix: Change of scattering coefficients by the shift of 
the one-dimensional (1D) potential}
Here we discuss the relation between the scattering matrix and the
translation of a 1D potential, and derive 
Equation~(\ref{eq:Delta-phi}). First, let us consider an arbitrary 1D
potential $V=V(x)$ that leads to the corresponding scattering matrix
\begin{equation}
 S = \left( \begin{array}{cc}
                r & t' \\
                t & r'
           \end{array}  \right) \;.
\end{equation}
Then the wave function $\Psi(x)$ for the electron initially injected from 
$x\rightarrow-\infty$ can be written as (at the asymptotic region)
\begin{equation}
 \Psi(x) = \left\{ \begin{array}{ll}
                     e^{ikx}+re^{-ikx} & (x\ll0) \\
                     te^{ikx} & (x\gg0) 
                   \end{array}
           \right. ,
 \label{eq:Psi-1d}
\end{equation}
where $k$ is the wave number of the electron.

Now assume that the potential is shifted by $\delta x$, that is, 
the potential is given as $V=V(\bar{x})$ where $\bar{x}\equiv x-\delta x$.
It is obvious that the wave function in the shifted potential 
$\bar{\Psi(\bar{x})}$ has the same form with $\Psi(x)$ given in 
Equation~(\ref{eq:Psi-1d}). That is,
\begin{equation}
 \bar{\Psi}(\bar{x}) = \left\{ \begin{array}{ll}
                     e^{ik\bar{x}}+re^{-ik\bar{x}} & (x\ll0) \\
                     te^{ik\bar{x}} & (x\gg0)
                   \end{array}
           \right. .
\end{equation}
Using the original coordinate $x$ instead of $\bar{x}$, one can find that
\begin{equation}
 \bar{\Psi}(x) = e^{-ik\delta x}
                 \left\{ \begin{array}{ll}
                     e^{ikx}+\bar{r}e^{-ikx} & (x\ll0) \\
                     \bar{t}e^{ikx} & (x\gg0)
                 \end{array}
           \right. ,
\end{equation}
where $\bar{t}=t$ and $\bar{r}=re^{2ik\delta x}$.

In other words, the transmission amplitude is invariant under
translation of the potential, but the reflection amplitude suffers
phase shift of $2k\delta x$. 
Applying the result derived here for discussion of phase-sensitive dephasing
in Section 3.4, we get Equation~(\ref{eq:Delta-phi}). 
\section*{References}

\end{document}